\documentclass[12pt]{iopart}

\usepackage[dvips]{graphicx}

\begin{document}

\title[The  waiting-time distribution of Liffe bond futures]
{The  waiting-time distribution of Liffe bond futures}

\author{M. Raberto\dag, 	     
E. Scalas\ddag, 		     
R. Gorenflo\S			     
\ and F. Mainardi\P		     
 \footnote[4]{To whom correspondence should be addressed
 (mainardi@bo.infn.it)}}

\address{\dag
	      Dipartimento di Ingegneria Biofisica ed
	      Elettronica, Universit\`a di Genova,
	      via Opera Pia 11a,  I--16145 Genova, Italy}

\address{\ddag
	     Dipartimento di Scienze e Tecnologie Avanzate,
	     Universit\`a del Piemonte Orientale,
	     via Cavour 84, I--15100 Alessandria, Italy,
	     and INFN Sezione di Torino, via P.Giuria 1,
	     I--10125 Torino, Italy}

\address{\$
	     Erstes Mathematisches Institut, Freie Universit\"at
	     Berlin, Arnimallee  3, D-14195 Berlin, Germany}

\address{\P
	     Dipartimento di Fisica, Universit\`a di Bologna and
	     INFN Sezione di Bologna,
	     via Irnerio 46, I--40126  Bologna, Italy}

\begin{abstract}

We apply the Continuous Time Random Walk (CTRW)
framework, introduced in finance by Scalas \etal \cite{SGM00},
to the analysis of the
probability distribution of time intervals between two consecutive trades
in the case of BTP futures prices traded at LIFFE in 1997.  Results
corroborate the validity of the CTRW approach for the description of the
temporal evolution of financial time series.

\end{abstract}



\maketitle

\section{Introduction}

In financial markets, the time interval between two consecutive
transactions (the so called {\it waiting time}) varies stochastically.
There have been various studies on the nature of the stochastic processes
generating the sequences of waiting times between two consecutive trades
\cite{Clark73,LM98}. In two recent papers, Scalas \etal \cite{SGM00} and
Mainardi \etal \cite{MRGS00} proposed the Continuous Time Random Walk
(CTRW) model \cite{MW65} as a phenomenological description of
tick-by-tick dynamics in financial markets. In the hydrodynamic limit, the
model yields a general scaling form \cite{SGM00,Iori00} for the
probability density function of finding the log-price $x$ at time $t$.
Scaling is a consequence of the equivalence between CTRWs and
fractional diffusion equations \cite{HA95,Hilfer00}.
The reader can also consult refs
\cite{BarkaiMetzlerKlafter00,MetzlerKlafter00}.
Here, as in \cite{MRGS00}, we will not focus our attention on scaling,
but on a different property: the waiting-time distribution.

This paper is divided as follows. Section 2 provides a brief summary of
the theoretical framework of CTRW. In Section 3 CTRW assumptions are
tested on market data. Conclusions are drawn in Section 4.

\section{\label{sec:theory}Theory summary}

Let us denote by $x$ the logarithm of an asset price. We assume that both
the log-price jumps $\xi_i = x(t_i) - x(t_{i-1})$ and the waiting times
between two consecutive trades $\tau_i =t_i - t_{i-1}$ are i.i.d.
stochastic variables. We also assume that these variables are
characterized by the two probability density functions: $\lambda(\xi)$ and
$\psi(\tau)$.

It turns out that
the evolution equation for $p(x,t)$, the probability of having the
log-price $x$ at time $t$, can be written as follows \cite{MRGS00}:

\begin{equation}
\int_{0}^{t} \Phi(t-t') \frac{\partial}{\partial t'} p(x,t')\, dt' =
- p(x,t) + \int_{-\infty}^{+\infty} \lambda(x-x') p(x',t)\,dx'\,,
\end{equation}
where $\Phi$ is a suitable memory kernel related in Laplace space to
$\Psi(\tau)$ - the probability that a given waiting 
interval is greater or equal to $\tau$ - by:

\begin{equation}
\widetilde{\Phi}(s) = \frac{\widetilde{\Psi}(s)}{1-s \widetilde{\Psi}
(s)}.
\end{equation}
In its turn, the probability $\Psi$ is given by:

\begin{equation}
\Psi(\tau) = \int_{\tau}^{\infty} \psi(t') \, dt'\,.
\end{equation}
The probability $\Psi(\tau)$ can be estimated from empirical data, and can
thus be used to test hypotheses on the memory kernel $\Phi$. In
particular, if $\Phi(t)$ exhibits a power-law time decay $\Phi \propto
t^{-\beta}$ with $0<\beta \le 1\,,$
 one can show that \cite{MRGS00}:

\begin{equation}
\Psi(t) \propto E_{\beta} (-t^{\beta}) \,,
\end{equation}
where $E_{\beta}$ is the Mittag-Leffler function of order $\beta$.
In the following, we shall discuss this particular hypothesis.

\section{Empirical analysis}

We examined the waiting-time distribution of BTP\footnote{BTP stands for
{\it Buoni del Tesoro Poliennali}, i.e., middle and long term Italian
Government bonds with fixed interest rates.} futures traded at
LIFFE\footnote{LIFFE stands for {\it London International Financial
Futures and Options Exchange}. It is a London-based derivative market} in
1997.  Throughout the year, there were four future contracts on BTP bonds,
depending on delivery dates: March, June, September, or December. We
considered two delivery dates: June (Figure 1) and September (Figure 2).
Usually, for a future with a certain maturity, transactions begin some
months before the delivery date. At the beginning, there are few trades a
day, but closer to the delivery there may be more than 1000 transactions a
day.  The total number of transactions was about 140,000 for the delivery
date of June and 170,000 for September.

Figure 1 shows a comparison between the function
$\Psi(\tau)$ estimated for the
empirical waiting times and two theoretical functions. The circles refer
to market data (delivery date of June 1997) and represent the probability
of a waiting time greater than the abscissa $\tau$. We have determined 494
values of $\Psi(\tau)$ for $\tau$ in a interval between 1 $s$ and about
50,000 $s$, neglecting the intervals of market closure.  The solid line is
a two-parameter fit obtained by using the Mittag-Leffler type function:

\begin{equation}
\Psi(\tau) =
 E_{\beta} [ -(\gamma \tau)^{\beta}]\,,
\end{equation}
where $\beta$ is the index of the  Mittag-Leffler
function and $\gamma$ is a time-scale factor, depending on the time unit.

We get an index $\beta=0.96$
and a scale factor $\gamma=1/13\,.$
The fit of Figure 1 has a reduced chi-square
$\widetilde{\chi}^{2} \simeq 0.2$
The chi-square values
have been computed considering all the 494 values.
The dash-dotted line represents the stretched exponential
function $\exp \{ -(\gamma \tau)^{\beta}/\Gamma (1+\beta)\}$, whereas
the dashed line is the power law function
$(\gamma \tau)^{-\beta}/\Gamma(1-\beta)$. The Mittag-Leffler function
interpolates between these two limiting behaviours:
the stretched exponential for small time intervals, and the power law
for large ones.

Figure 2 shows the results for the delivery date of September 1997. In
this case, we have 442 values of $\Psi(\tau)$ for $\tau$ in the interval
between 1 $s$ and about 50,000 $s$; as for data in Figure 1, we get
$\beta=0.96$ and $\gamma=1/13$; the reduced chi-square, computed
considering all the 442 values, is $\simeq 0.2$.

\begin{figure}[h]

\centering

\includegraphics[width=0.7\textwidth]{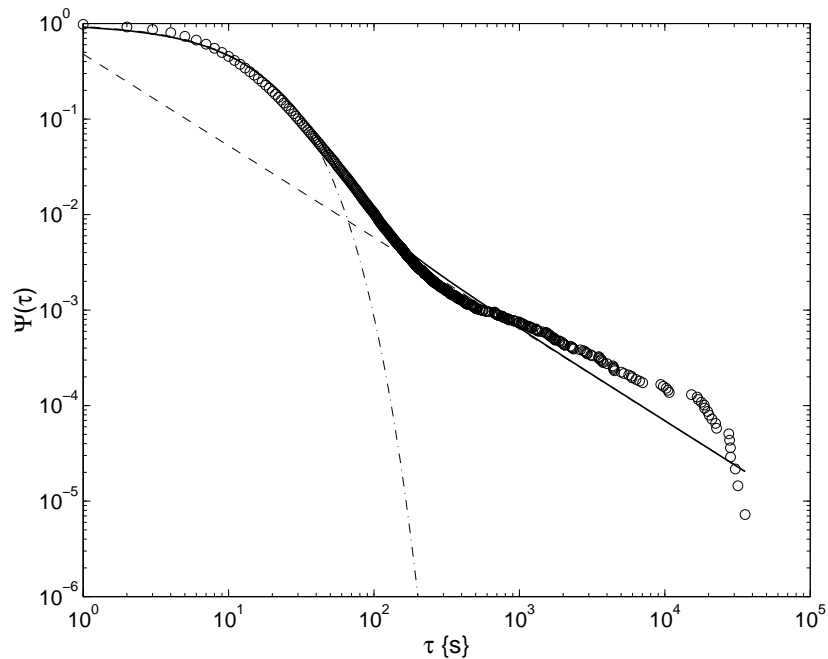}
\caption{Survival probability for BTP futures with delivery date:
June 1997. The Mittag-Leffler function (solid line) of index
$\beta = 0.96$ and scale factor $\gamma = 1 / 13$ is compared to
the stretched exponential (dash-dotted line) and the power
(dashed  line) functions.}

\end{figure}

\begin{figure}[h]
\centering

\includegraphics[width=0.7\textwidth]{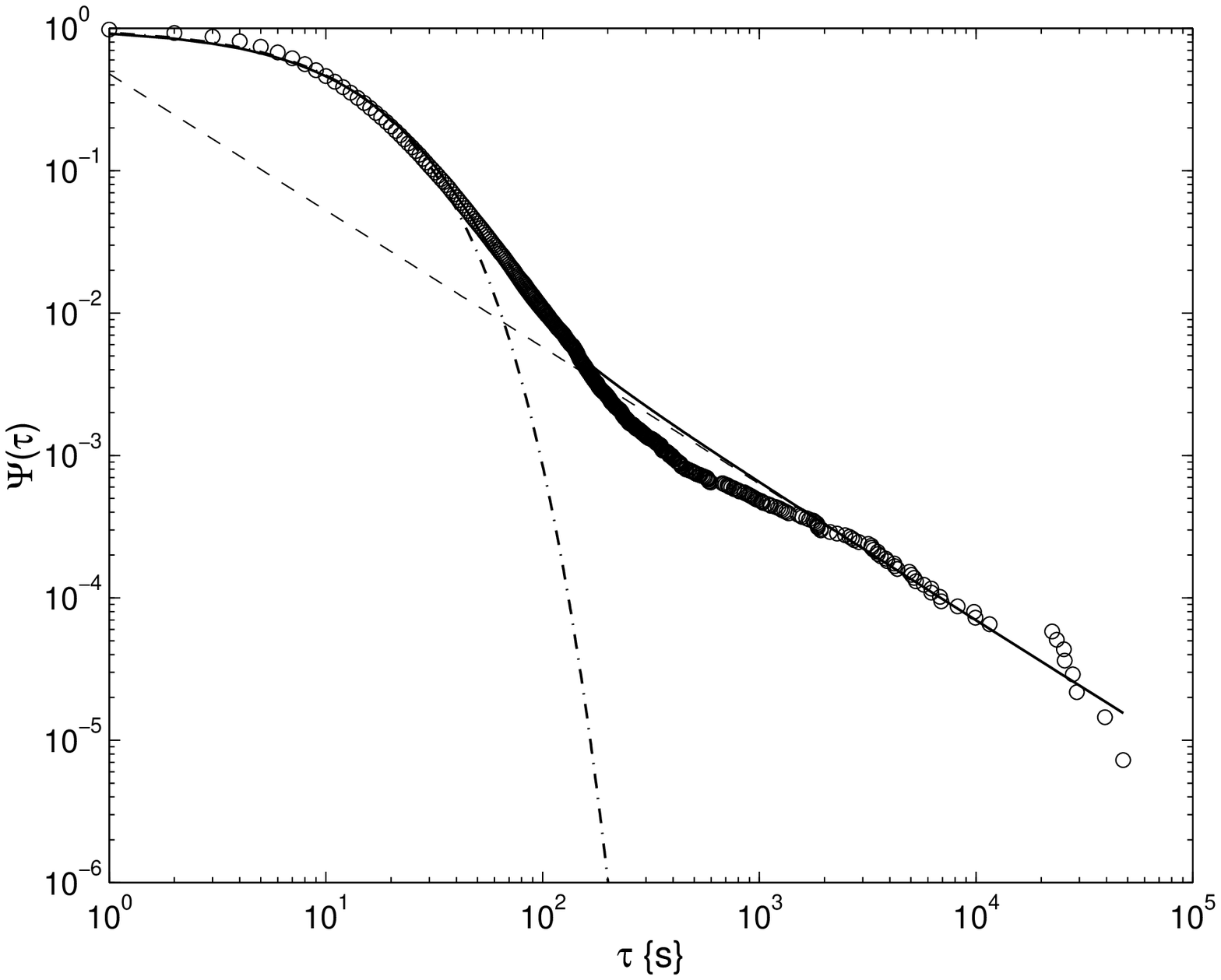}
\caption{Survival probability for BTP futures with delivery date:
September 1997. The Mittag-Leffler function (solid line) of index
$\beta = 0.96$ and scale factor $\gamma = 1 / 13$ is compared with
the stretched exponential (dash-dotted line) and the power (dashed line)
functions.}

\end{figure}

\section{Conclusions and outlook}

Figures 1,2 and the preliminary empirical analysis provided in
\cite{MRGS00} show a satisfactory agreement between the empirical
distributions of market data and theoretical predictions drawn from the
CTRW hypothesis with additional
assumptions. CTRW is thus likely to be a reasonable
phenomenological description of the tick-by-tick dynamics for a financial
market, as it also takes into account both the non-Markovian and the
non-local characters of the time evolution in financial time series.

The fitting procedure used in this paper differs from that in
\cite{MRGS00}. Indeed
we found the estimate of the function
$\Psi(\tau)$ and the fitting procedures far from
trivial. These points deserve a separate and thorough discussion which
will be the subject of a future paper.

\end{document}